# Beauty and art. Cognitive function, evolution and Mathematical Models of the Mind

**Abstract** – The paper discusses relationships between aesthetics theory and mathematical models of mind. Mathematical theory describes abilities for concepts, emotions, instincts, imagination, adaptation, learning, cognition, language, approximate hierarchy of the mind and evolution of these abilities. The knowledge instinct is the foundation of higher mental abilities and aesthetic emotions. Aesthetic emotions are present in every act of perception and cognition, and at the top of the mind hierarchy they become emotions of the beautiful. The learning ability is essential to everyday perception and cognition as well as to the historical development of understanding of the meaning of life. I discuss a controversy surrounding this issue. Conclusions based on cognitive and mathematical models confirm that judgments of taste are at once subjective and objective, and I discuss what it means. The paper relates cognitive and mathematical concepts to those of philosophy and aesthetics, from Plato to our days, clarifies cognitive mechanisms and functions of the beautiful, and resolves many difficulties of contemporary aesthetics.

## 1. Introduction

Based on novel ideas in the area of cognitive and mathematical theories of mind and new neuro-imaging data I would demonstrate relationships among concepts in cognition, mathematics of the mind, and aesthetics. I describe contemporary mathematical understanding of thinking processes, its conscious and unconscious aspects, mechanisms of concepts, instincts, emotions, cognition, language and their role in thinking. I discuss a possibility of understanding aesthetic emotions within mathematical theories of mind, the *understanding corresponding to the intuition of the beautiful* that have been escaping theoretical analysis in contemporary aesthetics and philosophy.

Relationships between mathematics and aesthetics (if any) are commonly thought of along the Plato's understanding of beauty as related to the world of Ideas and pure concepts; "remembrance" of this world enables us to enjoy beauty not in the objects of the world, but in relationships among objects. These relationships are manifested in mathematical constructs like symmetries, geometric shapes described by mathematical equations, or (more recently) fractal geometric objects. But these "mathematical" ideas of beautiful clearly lack the personal, the subjective element and therefore they do not address this necessary fundamental concern of aesthetics. Similarly, computer tools enabling computer-generated art do not address the aesthetics fundamentals.

Yet, the new mathematical models of mind are coming close to providing a new level of theoretical understanding of emotions, consciousness and unconscious, cognition and language, the individual and collective, and turning out to be closer than expected to the fundamental issues of philosophy and aesthetics. I will describe a complex play of the two factors of subjective and objective in the working of mind and relate it to more than two thousand year old debates in philosophy as well as to contemporary debates in aesthetics. In the next section I summarize results of mathematical theories of intelligence that are most relevant to aesthetics (while the exposition is mathematically correct, it is given at a conceptual level and no knowledge of mathematics is assumed throughout this paper; readers with knowledge and interests in mathematical details are invited to follow the references). Then, I will review the relationship of mathematical models to aesthetic concepts since Kant with detailed references to some of the issues discussed on the pages of leading contemporary journals in philosophical aesthetics[1] during the last decade. And, I will summarize fundamental mathematical arguments related to possibility or impossibility of the mathematical "formulae" of beauty.

Socrates asked all great Greek artists of his time about what was the beautiful, but nobody could give him an answer. Aristotle came close to conclusions of this paper, defining the beautiful as "unity in manifold." The word "aesthetics" was introduced by Baumgarten (1750); as a special ability to perceive the beautiful, separate from usual perception and cognition. Contemporary aesthetics began with Kant (1790), who based it on his theory of mind, and related to knowledge. Kant theory dominates contemporary aesthetics, however, because of certain difficulties discussed and explained in this paper, his theory has remained misunderstood. "The Cambridge companion to Kant" (Guyer 1992), which intends to explain Kant, could be understood only by experts, does not make Kant easier to read and does not clarify any of his thoughts beyond Kant (1790). This paper clarifies Kant's thoughts by comparing them to cognitive-mathematical theory developed in this paper.

In the following Section 2 I describe the mathematical theory of the mind including concepts, instincts, emotions. Section 2 analyzes a specific aspect of the mathematical theory of mind, the ability for learning and adaptation; a specific emotion involved in this ability, and presents arguments for this being the aesthetic emotion. Section 3 establishes detailed relationships between notions of the mathematical theory and those of Kantian theory of mind and aesthetics. Section 4 continues elaboration of the mathematical theory from "simple harmony" towards refined beauty. Section 5 relates the mathematical theory to current discussions of issues in aesthetics.

## 2. Intellect: concept, instincts, emotion

Analysis of mathematical theories of mind I begin with the processes most accessible to the consciousness. Those are the usual conscious perception of objects around us and comprehension of their usage. For example: "this is a chair, it is for sitting". It turned out to be difficult to create computers capable of such a simple perception-judgment. After fifty years of development of mathematical theories of intelligence, after spending billions of dollars, pounds, franks, marks, rubles, yen on this research, today we begin understanding the nature of the complexity of the problem. How to explain a complexity of the obvious? A word "chair" written on a paper is very different from the spoken sounds of the same word; a written and spoken "chair" is very different from a chair one sits on. In our brain there are inborn structures that were developed over hundreds of millions of years of evolution specifically to enable fast learning (in childhood) of combining into a single concept a spoken, written, drawn, and real chair. Let us note that the "real chair" is what is seen by our eyes, but also what is sensed as a "seat" by the sitting part of our body. Therefore "chair" is a bodily-sensed-spatio-thought concept.

The brain structures providing the foundation for learning concepts are called mental representations or *models* (mental models) of mind. They model the world around and the results of the modeling are *phenomena* perceived and understood by our mind. What are these models made of? What do they look like? What is the mathematics that could describe them? An original approach of the creators of artificial intelligence (in the 1950s and 60s) was based on an assumption that mental model-concepts contain "photo-likeness" of objects. An image of an object at an electronic eye "retina" was to be compared to mental models stored in the computer memory, resulting in decisions: similar-dissimilar. This early mathematical attempt to describe the working of the mind was based first, on the formal logic operating with binary variables (yes, no) and second, on the then known properties of the brain neurons, which seemed similar to early two-state transistors (0,1), 0=no, 1=yes. This approach failed: an object (say, chair) turned out to be different each time, a different view angle, lighting, surrounding objects, shape of legs or back – the computer could not "comprehend" that all of these are "chair".

The most difficult for mathematical models of object recognition turned out the fact that around an object of interest there always are other objects, most of no interest. Recognition, of even a single object, therefore required evaluating combinations of object. Combinations are mathematically "bad," because there are too many of them. When one looks in any directions there are very many objects, say, a chair, a table, a pattern on a floor, various scratches and spots on every object, etc., in various combinations. Even if one considers only 100 objects, not a large number, their combinations are $100^{100}$, a number larger than all elementary particle interactions in the entire life of the Universe. Clearly no mind or computer would ever be able to evaluate that many combinations, and select "useful" one for learning.

Today we know much more about complicated structures of the brain's neural networks (Grossberg 1982; Kosslyn 1980, 1994; Barsalou 1999). In neural modeling field theory, MFT, (Perlovsky 1988, 2001, 2006) mental representations, models, are vague and distributed, do not look like photos of objects, and not necessarily located in adjacent brain regions. They are described by fuzzy logic, which decisions are not similar to simple "yes-no". In the process of learning they "take a form" of a concrete object at its angle, illumination, etc. This process "from vague to crisp" is described by dynamic logic (DL), it reassembles objects and experiences from bits and pieces, and it was demonstrated in neuro-imaging experiments to actually take place in the human visual system during perception (Bar et al 2006, Perlovsky 2009c).

Vague models make up the content of a priori, inborn structures of mind, providing the foundation for perception and cognition (Perlovsky & McManus 1991; Barsalou 1999, Perlovsky 1994a, 1997a, 2002a, 2007c,d,f, 2008a, 2009c, 2010d). These structures are not accessible to consciousness directly. Jung called some of these structures archetypes of mind (Jung 1921). At the moment of birth, and possibly to some extent, even before the birth begins the process of learning-adaptation of these a mental structures to the surrounding world. On the basis of inborn mental structures, the models of concrete objects and situations are learned. Models related to vision, hearing, olfaction, touch are created for perception and correspond to objects in the outer world. In other words, mental models of perception correspond to object-structures in the sensory signals generated by retina and other sensory organs.

Based on knowledge of neural mechanisms, it is possible to demonstrate to oneself in 3-second experiment, that models-representations are vague and unconscious. Look at an object in front of you, then close eyes and imagine this object with closed eyes. The imagined object is not as clear and crisp as it is with opened eyes. The imagination is vague. During recent decades we learned that imaginations are produced by top-down signals from concept-models to the visual cortex. Therefore we conclude that models are vague. In interaction with sensory data these vague representations become crisp and clear. The vague models are also less conscious.

Cognition and understanding of abstract concepts of mind are similar to "simple" sensory perception of objects (Perlovsky 1997, 2002a, 2004, 2005, 2006a, 2007a,d,e, 2008a). In "simple" sensory *perception*, phenomena-objects are formed from sensory neural signals based on mental models of objects. Similarly, new concepts in abstract *understanding* processes are formed on the basis of more complex models from concepts understood earlier. This process continues up and up the hierarchy of the concepts of mind toward the more general and abstract concepts of complex situations and relationships. In MFT-DL, mathematical models describe concepts of simple objects as well as abstract concepts of complicated situations and relations. Whereas *object*-models represent geometric, color, and other sensory properties of objects, the models of *abstract concept models* represent relations among objects. Mental models of abstract concepts as well as mental object-models are distributed and vague (Perlovsky 2001, 2006a, 2009c Bar et al 2006, Barsalou 1999). Models higher up in the mind hierarchy are built on top of several layers of vague and less conscious models, correspondingly, they are even more vague and less conscious. In the process of perception and cognition, they take a structure of concrete concepts, approximately obeying usual logic. In such a form of concrete objects, situations, and relations they become accessible to consciousness and are perceived as objects and situations in the world, or as thought objects. Let me emphasize that the DL processes of perception and cognition, as well as vague representations, are not conscious; only the final states of DL processes, approximately logical states of mind, are accessible to consciousness.

(Let us illustrate the above with simple examples. We perceive a concrete chair on the basis of a vague and distributed representation of chair, containing *vague* notions of a seat, back, and legs, which might be stored in various brain areas. In the moment of perception, these pieces of past experiences are assembled into an image of the concrete chair with *concrete* seat, back, and four legs. A more complicated situation is understanding, say, that you are in a concert hall. Your understanding is based on a mental model that includes a scene, a big room, and many chairs arranged in rows.)

What are the relationships between concepts and basic needs of an organism? (For example, how to explain that a hungry person "sees food all around?") The basic needs of an organism, e.g. eating, are indicated by *instincts*. Instincts are like internal sensors measuring vital organismic parameters, and generating signals in neural networks indicating fundamental needs. Mechanisms of concepts have been selected by evolution to satisfy instinctual needs. Connection of instincts and concepts is accomplished by emotions. In a usual conversation, "emotions" refer to a special type of behavior: agitation, higher voice pitch, bright eyes, there are several mechanisms of emotions. In this paper I follow instinctual-emotional theory emotions developed by Grossberg and Levine (1987). Every instinct generates evaluative emotional signals indicating satisfaction or dissatisfaction of this instinct. These signals affect the process of comparing concept-models to objects – the process called *judgment* by Kant (1790). Emotions are evaluations "good-bad". Evaluations not according to concepts of good and bad, but direct instinctive evaluations. An emotion evaluates a degree to which a phenomenon (object or situation) satisfies our instinctual needs. Most of emotions originate in ancient parts of the brain, relating us to primates and even to lower animals. Now we consider higher emotions originating in cortex and related to knowledge.

### 3. The Knowledge Instinct and aesthetic emotion

*...All was mixed at the beginning;*
*but mind appeared and created order.*
*Anaxagoras*

Imagine for a moment that you cannot see clearly one object from another, do not understand their relations and purpose; sounds merge in noise, where you cannot discern their sources or directions, there are no clear thoughts in consciousness and will is incapable of concentrating in a concrete desire. A human cannot survive in such a horrible state.

An adequate perception of the surrounding world, an understanding of relationships in the world, and an ability to concentrate will are so important for survival, that it can even outweigh other "bodily" needs. Better said, an adequate perception, orientation, understanding, and concentration of will are conditions of satisfaction of bodily instincts. The need for "understanding" is so important that there is an unconditional, inborn mechanism, an *instinct*, driving this need; driving it independently from satisfaction or dissatisfaction of other instincts.

What is the mechanism of this instinct and how do we perceive it in our consciousness?

Perception and cognition consist in adapting inborn, culturally-received, and individually learned mental models to concrete conditions around us. The faculty of judgment associates an individual object with the general rule, as discerned by Kant (1790, Intro. IV), and also adapts, modifies, the models in this process. Mental models of mind have to be brought in correspondence with the world. Correspondence between the models and the world is knowledge. That is why the instinct driving the mechanism of adaptation is called the *knowledge instinct* (KI). This same instinct makes little kids, cubs, piglets jump around and play fight, their inborn models of behavior must adapt to their body weights, objects, and animals around them long before the instincts of hunger and fear will demand their usage for direct aims of survival; neural mechanisms involved in KI were discussed in (Levine & Perlovsky 2008). Kiddy behavior just makes the work of KI more observable; to varying degrees, this instinct continues acting all our life. All the time we are bringing our internal models into correspondence with the surrounding world.

Mathematically KI is modeled by maximization of a similarity measure between models and sensory perception. More generally, we should consider a hierarchical mind structure; representations-models at a higher level aim at unifying subsets at lower levels; top-down signals coming from higher levels interact with bottom-up signals coming from lower levels, and KI drives this interaction toward increased similarity at every level.

In biology and psychology KI has been discussed as curiosity, cognitive dissonance, or a need for knowledge since the 1950s (Harlow 1950; Festinger 1957; Cacioppo & Petty 1982). In computational intelligence it is ubiquitous, every mathematical learning procedure, algorithm, or neural network maximizes some similarity measure. As every instinct, KI involves knowledge-related emotions evaluating satisfaction of this drive for knowledge (Grossberg & Levine 1987; Perlovsky 2001, 2000c, 2006a,b; 2007c; Perlovsky, Bonniot-Cabanac, & Cabanac 2010).

How can we understand the nature of emotions corresponding to KI? Let us direct our inner gaze on our consciousness: are emotions satisfying or dissatisfying the learning instinct accessible to consciousness and what do they feel like? Let us return to the beginning of this section and imagine once more a situation of utmost disorientation; objects and their relations are indiscernible, and those that are behave unpredictably: doors cannot be opened, water does not pour, the teeth cannot bite the apple, the will does not concentrate nor direct our actions, and nothing makes sense – a terror takes hold. On the opposite, when surrounding objects and people behave according to the ideal expectations and needs, and actions reach clear aims we feel pleasant harmony, we like it. A feeling of harmony is a correspondence between the inner concept-models and the outer world. Emotions of pleasure and terror related to satisfaction or dissatisfaction of the learning instinct are *aesthetic emotions*. As discussed later, aesthetic emotions are the foundation of our higher spiritual abilities; emotions related to knowledge have been called aesthetic since Kant. Experimental investigation of these emotions have been initiated in (Perlovsky, Bonniot-Cabanac, Cabanac 2010), and cortical mechanisms involved in aesthetic emotions were discussed in (Levine & Perlovsky 2008).

## 4. Cognitive-mathematical theory of knowledge and Kantian aesthetics

*Is beautiful... flowing and changing... or always such as it is?*
*Socrates*

Kant founded aesthetics on his theory of perception and cognition, based on processes of interaction between our psyche and the surrounding world. I briefly recollect here some main ideas of Kant relating them in detail to mathematical concepts of mind. As well known, by discovering the a priori inborn nature of human psyche as the basis of cognition Kant overturned the historical course of the human thought and redirected the philosophical analysis of relationships of the human and the world, objective and subjective, particular and universal. He discovered that our perception and cognition is related to the world of phenomena and not to the outer world of things-in-themselves. Phenomena are created by our mind in the process of interaction with the surrounding world, with things-in-themselves. A specific ability for creating phenomena (Understanding) is based on a priori structure of our mind: every phenomenon is created on the basis of the corresponding structure. (In DL, phenomena are final states of models). We are capable of orientation in the world due to (and to the extent of) our universal concepts, mental models, being adequate or similar (to some extent) to the corresponding individual objects-in-themselves.

Yet, the concepts and objects are different in their nature. So, how is it possible to establish correspondence between concepts and objects? By using a special a priori (inborn) ability, established Kant, the ability for judgment; judgment of which concept-model corresponds to which (individual) object or event (1790, Intro. IV). In the NMF-DL it is modeled by measure of the correspondence or similarity between models and sensory data (Perlovsky 2001, 2006a). Kant has further demonstrated that the ability for judgment is the foundation for the ability to perceive beauty.

Our reason prescribes *causality* to the world of phenomena, as its basis, its fundamental a priori law: in the world of phenomena, one follows another and there is no room for freedom. But in the area of morals governing our will, reason prescribes *freedom*, as a fundamental a priori law of behavior (1790, Intro. IX). Everyone considers oneself free, at least, in thought. Freedom and causality seem to be irreconcilable in principle; practically, however, this irreconcilable contradiction is resolved every day by everyone in many minute situations. For example, coming home from an office she eats dinner. First, she behaves according to the law of causality, according to which the food is needed to support life. Second, she does it completely free, according to her free subjective choice. The contradiction has been reconciled due to the ability of judgment to make a correct choice among many alternatives in correspondence with the general concepts of objects (dinner and body, which are subjects to the law of causality) and concepts of behavior (free desire to eat). Of course, life is far from being that simple all the time, but it is important for Kant that such ability for judgment exists in principle.

Let us note that in the above example a person receives a pleasure, and lets direct our attention to its dualistic nature: first, a hungry person enjoys eating, and second, a human made a correct choice in a situation. This "second" is a little example illustrating the great discovery by Kant: the correspondence between concepts and world established by judgment ability brings pleasant satisfaction. Kant comes to *the fundamental a priori law governing judgment ability*, it is purposiveness. It is not the world of phenomena, which is purposeful according to some a priori law. No. Judgment ability is *purposeful*. However, we perceive this situation as if objects-in-themselves (world and behavior) were purposive: a human being freely desires, acts according to reason, and receives pleasure and benefits at the same time.

Pleasure brought by judgment ability could appear without satisfaction of "lower" bodily needs related to hunger or fear. Sometimes we enjoy pleasure without expecting any specific benefit. This way, according to Kant, we experience beauty and spiritually sublime (1790, Intro. IX). Beauty is *aimless purposiveness* perceived in an object, particularly, in objects of art. Spiritually sublime is aimless purposiveness perceived by consciousness in a will, models of behavior and actions, either, our own, those of other people, in objects or cultural symbols. "Aimless" means here directed *not* at the satisfaction of "lower" materialistic instincts and needs.

In the mathematical theory, "pure" *purposiveness* of judgment is described by DL as maximization of likelihood (or information), in other words, the drive to *improve the correspondence between concepts (models) and the world*. It is definitely not aimless; it is directed at a goal, which is extremely important for survival. However, this goal is not directly aimed at bodily "low" materialistic satisfaction, but to the satisfaction of KI. Kant saved no effort in emphasizing that his "aimless purposiveness" (sometimes translated as "purposiveness without purpose") is purposive in the highest spiritual sense (1790, §§1-22). Yet, without a notion of KI the exact meaning of this purposiveness has remained unclear (Kant has discussed this limitation of his theory in 1790, §2). Debates, with much misinterpretations and their rebuttals, around this central Kantian idea have begun since Schiller and continue to this very day (Schiller 1895, Spencer 1891, Guyau 1884, Perlovsky 2000, multiple current references are discussed later).

It is interesting to compare the Kantian theory of mind to certain ideas of Buddhism. Buddhism penetrates into the depths of perception and cognition processes by recognizing that phenomena are not identical to things-in-themselves. Emphasizing this, Buddhism refers to the world of phenomena as "Maya", the meaningless deception. A fundamental Buddhist notion of "emptiness" points out that the value of any object for satisfying the "lower" bodily instincts is much less than its value for satisfying higher needs, KI. Consciousness of bodhisattva, writes Dalai Lama wonders at perception of emptiness in any object (Dalai Lama XIV 1993). Any object is first of all a phenomenon accessible to cognition. Bodhisattva's consciousness is directed by KI. The concentration on "emptiness," therefore does not mean emotional emptiness, but the opposite, the fullness with highest emotions related to KI, beauty and spiritually sublime (Perlovsky 2001). Similarly to Buddhist "emptiness," Kantian "aimless purposiveness" is directed at the satisfaction of KI and is purposiveness in the highest sense. It is also interesting to note that Aristotle came close to this definition of beauty as "unity in manifold", for our transcendental idea of the world in its totality is of the purpose (end) without aim.

After formulating the idea of beauty as purposiveness without purpose, Kant moves on to analytics, deductions, and dialectic, giving a number of examples to clarify his thoughts and explaining why the beauty is both objective and subjective; yet, according to von Hartmann "no other thought caused so much difficulty to Kant and turned out so controversial." And his own thoughts on fine art (1790, §51-54) contradict his fundamental theory of the aesthetic pleasure.

Pointing out to this complex question of the dialectic of subjective and objective in the beautiful, Kant's system did not resolve it. If this insightful reformer of philosophical thought did not break up the enchanted circle of objective and subjective, what is the reason? And where could we look for the basis to break out from this circle? DL and neuroimaging data penetrated into the mystery of conscious and unconscious, turned around understanding of the role of logical thoughts in philosophy and psychology, and led to conscious contemplation of the *issue of the historic dialectic of the beautiful*, the change of the notion of beautiful in the historic change of consciousness. Let us look again at the classical philosophical question of the relationship between apriority (initially given, inborn abilities) and adaptivity (learning), and concentrate on the changes in understanding of these issues brought about by the development of science (Perlovsky 1998c, 2010h).

Apriority was understood in all classical philosophy, beginning from ancient Greeks, as absolute and eternal constancy, immutability: a priori concepts of mind never change, their origin is in the other world, beyond our perception, and they are the basis for all the objective. There was an impenetrable barrier between the apriority, as an immutable transcendental givenness, and adaptivity, as variability, changeability and the basis for individual differences. This opposition and irreconcilability of apriority and adaptivity was inherited by Kant, striving to discover the eternal and immutable, a priori laws of mind analogous to the eternal and immutable laws of nature discovered by Newton[2]. Yet, science, invading the domain of philosophy repeatedly anew discovers the play of the a priori and adaptive, objective and subjective, and overcomes the barrier between apriority and adaptivity. Attempting to discover the origins of Universe, physicists are coming to a conclusion that even the fundamental laws of nature are changing with time. The meaning of the apriority of mental abilities changes: the a priori for an individual (like apriority of space and time) changes in history and is transferred (between people and generations) through language and cultural concepts. Adaptivity-learning of new concepts of mind created in culture (that is in the language and throughout life) is based on changeability of the neuro-physiological apparatus of a priori concepts. A new notion of adaptation of inborn concepts and the idea of KI connecting apriority and adaptivity leads to a new way of understanding the contradictions that were brought into the philosophical consciousness in the post-Kantian development of psychology and aesthetics. I would like to emphasize that it is not an accidental lapse of analytic sharpness that prevented Kant from understanding beauty as an emerging and historically changing property. This could have only been accomplished after the third Critique was completed, and it requires re-writing of all three Critiques as parts of a dynamic system involving individual mind and culture in their historic evolution. Several times Kant came close to formulating this problem (1790, §75). It is not a "magic of mathematics", but the gradual historic change in the understanding of the nature of the a priori and the adaptive that leads to this new insight.

## 5. Hierarchy of values and beauty

*Everything must be conscious to be beautiful*
*Socrates*

Let us repeat that aesthetic emotions appearing in the NMF-DL theory of mind are related to KI (mathematical measures of correspondence between concept-models and the world, between higher and lower representations), and these emotions are experienced independently of other instincts. They are not related directly to hunger or satiation, fear or comfort, sexual procreation. A feel of harmony (between concept-models and the world) is independent, to some extent, from concrete "lower" bodily needs. The aesthetic need is "spiritually high".

Harmony is just a first step toward the notion of beauty. Can aesthetics based on harmony of internal concepts-models and outside objects explain a feel of beautiful when one looks at a flower? A flower seems to be an insignificant object from the point of view of the survival of the human species. Why evolution could lead to ability for feeling a special harmony between our concept-models and a flower? Object-flower is simple, but the quest for beauty is complex; to find an explanation for a deeper meaning of harmony I turn now to the hierarchy of the organization of mind.

Harmony so far was meant as an "object"-harmony: correspondence between models of perception and objects in the world. This is a harmony at a lower level in the hierarchy of concepts of perception and cognition. At this level concept-models correspond to individual objects. "Higher up" in this hierarchy are more abstract and more universal concept-models of mind, modeling relationships among objects, and on to the models of situations, relations among situations, and, at the top of this "pyramid" there are concept-models of the meaning of our existence. Without interacting with objects and events in the world, these concept-models are vague and uncertain; they are the bases for learning, and inaccessible to consciousness. In the process of learning-cognition (the DL process, Perlovsky 2006a, the simulator process of perceptual symbol system, Barsalou 1999), they attain their "end-forms", more concrete and more accessible to consciousness. Evolution of vague unconscious forms into crisp conscious forms occurs in the interaction of concept-models and sensor signals (or higher level representations and lower level bottom-up signals).

Let us consider in some details contents of abstract models near the top of the mind hierarchy. Models at every higher level appeared in the process of biological and cultural evolution with a purpose to unify multiple lower level models. Recognizing a table, silverware, chairs as individual objects will only bring our understanding that far. KI drives our mind to understand the purpose of these objects, which we understand due to a higher level model of a dining room. Models at the top of the mind hierarchy attempt to unify our entire life experience, to understand it as something purposeful. Even so these models are vague and mostly unconscious, we "perceive" them as related to the purpose of our life. This "perception" is vague and unconscious we feel it as a mixture of conceptual and emotional contents. Contemplating an object outside self stimulates the process of adaptation; in the first moment, model-objects are adapted, providing for perception; in the next moments model-relationships and model-situations are adapted, and this process continues up and up the hierarchy of models, reaching sometimes the highest models of the meaning of life (Levine & Perlovsky 2008; Perlovsky 2010h).

A contemplation of an object, therefore, could stimulate adaptation of abstract complex models, bringing a deep satisfaction of the instinct for learning. Aesthetic emotion in such a case is perceived by us as a presence of beauty. A flower seems beautiful to us because we see in it a purpose – the process of our internal adaptation reaches concept-models of the meaning and purpose of existence. Possibly, it is related to the fact that a flower is full of biological meaning. Flowers appear relatively late in evolution; ferns do not yet have flowers, it appears first at horsetails belonging to a botanic classification of vasculiform cryptogamous. A flower attracts bees by appearance and scent. Its geometry is exactly "calculated" so that a bee's head touches stamens exactly by the same spot that later touches pistils. Leaves and branches of a plant can grow larger or smaller depending on rains, sun, temperature, but the geometry of a flower does not depend on these accidentals. After pollination the flower dies and a fruit appears in its place. Even being not a botanist and knowing nothing about the biological function of a flower, a human is capable of feeling that this little object is full of meaning. This fullness of meaning in a simple object gives us a possibility to feel meaning in nature in general, and in particular, the meaningfulness of our existence. A flower hints to our unconscious about their joint evolution lasting billions of years. Knowing of the biological function of a flower, procreation, one may conclude that the meaning of a flower is related to a finite goal. However, being a part of the natural evolution, including human, a flower as if sets ajar a possibility of a meaning aimed beyond a concrete aim, what Kant called *purposiveness without purpose*. The beauty of a flower is a perceived possibility of improving ourselves, improving the inner concept-models of our purposiveness toward the aim that is hidden from us as of yet. A flower example illustrates both, a possibility of change of aesthetic criteria and slowness of this process: for a human being who learned botany a flower beauty may lose a part of its mystery (aesthetic fascination may be shifted toward scientific analysis), yet the biological nature of our emotions changes very slow, and flowers will continue bringing us aesthetic joy for thousands of years.

Emotions of the beautiful is an aesthetic emotion related to satisfaction of KI at the top levels of the hierarchy of mind. As discussed, every mental representation has a purpose of unifying multiple lower-level. Representations at the top have a purpose of unifying the entire life experience. It is vague and not conscious, its conceptual and emotional contents are not differentiated. Therefore, the concept of purpose of life and the emotion of the beautiful cannot be made as clear as an object in front of our eyes. Even if an object or event can make us more certain that the purpose of life exist, we feel the presence of the beautiful.

However, sometimes we think we know what is the purpose of our life and we can discuss in details attributes of the beautiful. To reconcile the contradiction between vagueness and certainty, we need to consider interaction between cognition and language (Perlovsky 2004, 2006a, 2007b,e,g, 2009a, 2010g; Fontanari et al 2007, 2008a; Tikhanov et al 2006). Language and cognition interact through the dual model: each mental representation has two parts, language and cognitive. Language representations are learned from surrounding language, their learning does not need life experience, therefore they can be learned early in life. By five years kids can talk about the entire content of culture. The entire hierarchy of their language representations is crisp and conscious. But they cannot act as adults, their cognitive representations are vague. Throughout life cognitive representations are learned and become more crisp and consciousness guided by language and grounded in life experience. The higher in the hierarchy, the more vague and less conscious are cognitive models, whereas language models remain crisp and conscious. We can talk crisply and consciously about contents of top representations using language, which accumulates the cultural wisdom. And we may remain unconscious about vagueness of corresponding cognitive contents. This is why sometimes people talk clearly, crisply, and knowledgably about contents of the top models. This may help us improve contents of corresponding cognitive models, which forever remain vague and unconscious. Yet KI drives us to make these contents crisper and more conscious. When we achieve even a small step in this direction, we feel emotion of the beautiful.

Aesthetic sensitivity of a human could be directed not just outward, at outer objects (flowers), but also inside self, at the inner concepts and intuitions. This is the nature of aesthetic emotion in philosophy and science: harmony is a correspondence of general abstract concept-models (scientific theories) to particular, concrete concept-models which meaning is clarified in this correspondence. E.g., the concept of "mass" attained an exact meaning in the second law of Newton concerning the relationship between mass, force and acceleration. "Mass" attained even deeper meaning in the general theory of relativity, which related mass to energy and curvature of space-time. This internal connectedness of phenomena, a meaning that a scientific theory ascribes to the nature is perceived by scientists as beauty of a scientific theory. Beauty of an internal object in the case of science is a possibility to improve our internal most general and abstract concept-models of meaning and all-encompassing scientific theories, which we feel as related to a concrete thought, hypothesis, or experimental data (internal or external object). This is why founders of contemporary science, Einstein, Poincare, Bohr spoke about beauty of a scientific theory as the first criterion of its validity.

In art, creativity has been historically directed predominantly at an outer object, an art object. However, the role of inner object increases, especially beginning from XIX c., the thoughts of an artist or connoisseur stimulated by the work of art are becoming more independent from and of more value than the object of art itself. That is, the aesthetic emotion is directed not at the piece of art, but increasingly at the inner object. Despite increased specialization, the nature of creative process brings closer the artist, writer, poet, philosopher, and scientist. Beauty of an internal object as well as of an external object is a possibility to improve our internal concept-models of meaning of our life and existence that we feel when contemplating an object of art.

## 6. Aesthetic and anti-aesthetic in art, cognitive-mathematical model of mind, and contemporary aesthetics

Concentration on inner object was an inseparable part of enjoying, perceiving, and understanding art for millennia. The art of tragedy cannot be understood in any other way, and the Aristotelian explanation of the catharsis as a mechanism of aesthetic perception of tragedy concentrated on the inner object. Beginning from XIX c. this process has fast accelerated creating tensions, misunderstandings, and arguments about the role of "non-beautiful" in art. The process of accelerating art toward inner object is neither simple nor straightforward and in the process of developing new expressive means, artists often placed beauty at the backseat, trying first to cope with the basics of the aesthetic perception. This caused much discussion and controversy in aesthetics as the very notions of art and art object came under scrutiny. The complexity of the evolution of art in the 20$^{th}$ century brought many to a conclusion that art is impossible to define (Weitz 1956).

In this section I concentrate on relating the mathematical-cognitive analysis to several issues currently discussed in theoretical aesthetics. The mathematical theory of mind touching upon the foundations of aesthetics potentially offers relevant comments to good many current discussions in journals on philosophic aesthetics and art criticism. I limit my comments to the following four: the notion of aesthetic attitude and the surrounding discussion of what constitute the specifics of aesthetic; the dynamics of conscious and unconscious in creativity; conditions leading to modernity and post-modernity; and the "anti-aesthetic" in art. Even this limited selection of issues encompasses hundreds of publications making it impossible to comment on all of the aspects of the discussed problems (just to list a few, Gaut 2000, Budd 2001, Carroll 2002, Levinson 2002, Matravers 2003, Livingston 2003, Dickie 2004, Davies 2004, Gaut 2005, Carroll 2006, Matravers 2007, Scruton 2007); my purpose here is not a comprehensive review, but an attempt to establish a relationship between the cognitive-mathematical theory and aesthetics.

In addition to devastating the notion of aesthetics as a science, an idea of art non-definability ran contrary to the tradition dated at least to Kant. Countering the idea of "non-definability" of art, and to define art and art perception as a specific type of experience Stolnitz (1960) introduced a notion of aesthetic attitude. He emphasized that the attention is selective and directed to achieving specific goals, however, it is never exclusively "practical" and sometimes directed at the enjoyment of the "looks or sounds or feels". And he defined "the aesthetic attitude as disinterested and sympathetic attention to and contemplation of any object of awareness whatever, for its own sake alone". Comparing this definition with the aesthetic emotion as it appears in the MFT-DL theory of mind, one notices on the one hand, significant affinity (aesthetic emotions being specific and fundamentally different from other emotions, being "free" from immediate bodily practical interests) and on the other, uncertainty and limitation concerning "disinterested-ness" and "sympathy". For, let me repeat again, aesthetic is "interested", but in a higher sense, and not limited to "sympathetic" feelings but open to positive and negative emotions (as well as to complicated superpositions and interactions of "positive" and "negative" at once at different levels of the hierarchy of mind). These deficiencies of Stolnitz' definition are similar to those of Kant. Both Kant and Stolnitz explained in detail what they meant by "disinterested-ness", yet concise and clear *positive* definition (what it *is* rather than what it is *not*) was missing because complete scientific understanding did not exist; this left "the aesthetic attitude" open to misunderstandings similar to those faced by Kant's theory (Danto, 1996).

Criticism of the notion of aesthetic attitude is related to complexity of artistic experience in XX c. and goes to the roots of difficulties faced by aesthetic theory. Considering aesthetic attitude as a form of attention, Dickie concluded that such a notion is self-contradictory and cannot be satisfactory defined (Dickie 1964). Dickie's criticism of aesthetic attitude demonstrates complexity of the issues involved. Scientific analysis of the working of mind can help in sorting out the subtleties, which turned out to be important for aesthetic theory. Identification of aesthetic emotion with a special form of attention is wrong. (While Stolnitz himself used the word "attention", for him this was only "a point of departure". However, when Dickie criticized "attention" he departed in a very different direction from Stolnitz'). There are several mechanisms of attention in mind; all of them are different from emotions in principle. Attention is an amount of the mind resources allocated to a phenomenon based on emotions associated with this phenomenon. Emotions are evaluative signals and the corresponding feelings related to a phenomenon's ability to satisfy or dissatisfy a specific instinct. And, aesthetic emotions, to repeat it once more, are related to the phenomenon satisfying or dissatisfying our cognitive ability or more specifically KI.

The difference discussed above might seem subtle, but I am concentrating on analyzing it because it is related to a longstanding fundamental error about the nature of mind shared by philosophers and mathematicians, and leading to much confusion in theoretical aesthetics, the error of considering mind as a logical system. Its origins can be traced to Plato's conception of mind (IV BCE a) (even before Aristotle created logic). Aristotle (IV BCE) on the one hand founded the science of logic, on the other, he was specific about logic not being a theory of mind. He related logic to language, emphasized that language does not obey the fundamental law of logic ("the crispiness" given by "the law of excluded middle"), and warned not to make logical definitions more exact than language. These subtlety of Aristotelian thoughts were not understood, and later have been seen as the contradiction between "crispiness" of logic and "fuzziness" of language; Aristotle developed a theory of mind (forms) similar to DL (Perlovsky 2001, 2007c, 2010c). It drew the attention of Boole (1847) and other founders of formal logic, including Frege and Russell, and the perceived contradiction was resolved in the 19$^{th}$ c. by eliminating (from logic) the fuzziness of language. The consequences were unexpectedly "disastrous" for logic. The internal contradiction and inconsistency of "formal" or "crisp" logic was proven by Gödel (1933/1986). Logic turned out to be neither as omnipotent nor as "logical" as was expected. It follows that the mind does not work according to logic (Perlovsky 2006a,b, 2007c,g, 2008a, 2010c,d,g, Perlovsky & Ilin 2010a,b). Nevertheless, this implication of the Gödel theory was not understood by many philosophers and mathematicians, the erroneous idea of mind operating as a logical system found its support, among many others, in Wittgenstein (1921), Putnam (1988), and Newell (1983).

Today the idea of logical mind is discredited among mathematicians due to the fact that more than forty years of attempts to base mathematical intelligence on logic failed. Yet, "deification of logic," equating mathematics and science with logic is still a powerful myth in collective consciousness of today. The reason is that, as discussed, vague states and processes in the mind are not accessible to consciousness; consciousness can only access logic-like states;

therefore mind appears logic-like to consciousness. When considering mind as a system operating with logical rules according to the laws of logic, there is no fundamental difference between phenomena, emotions, and attention, all of these are sorts of logical rules, that is concepts. Thus, the error of Dickie is not just in missing some barely relevant distinction, but is related to an incorrect conception of mind that was dominant among mathematicians from 19$^{th}$ c. till 1970s (under the names "logical formalism," "Artificial Intelligence"), and among many leading philosophers of the 20$^{th}$ c., being directly related to Plato's conception of mind. Understanding mind as a logical system eliminates a distinction between concepts and emotions and inevitably leads to defining aesthetic contents according to a concept (not emotion).

Another aspect of Dickie's criticism has been directed at "disinterested-ness" of the aesthetic, which he has understood as irrelevancy to real life and human feelings. This misinterpretation is widespread. Conceptual difficulty in comprehending "disinterested purposiveness" led many to forget the "purposiveness" in Kantian theory and to emphasize the "disinterested-ness." Dickie's criticism helped to identify fallacious misunderstandings of this idea, especially when it was directed at the popular notion of "psychical distancing," which carried the aesthetic from "disinterested-ness" further into the realm of "irrelevancy," and it is this interpretation that drew his sharpest criticism. The tradition of interpreting Kant's theory of aesthetic, beauty, and "free play" as enjoyment of art for art's sake, that is irrelevant for "real life," has a long history descending from Schiller (1895). Discussions of "art for art's sake" pervade much of the aesthetics literature to this very day, it continues in Scruton (1974), Guyer (1997), Zangwill (1995), and Kemp (1999) (to mention just a few). MFT-DL explains "free play" as the working of the mechanism of KI, which is "free" in the very specific Kantian sense: free from "lower" bodily instincts. The "negative" definition of Kant (of what "disinterested" is not), as I have discussed, proved insufficient and led to much misunderstanding throughout history. And, let me repeat again, the "positive" definition of the specific nature of aesthetic *interest* became possible due to the theory of KI, the mechanisms of learning and adaptation of mental concept-models, the dynamic nature of mind, of the "faculty of understanding."

An opposite direction of misunderstanding of the "disinterested-ness" is in denying it altogether. Aiken (1998) attempted to explain art "ethologically", similar to other types of instinctive behavior. She asked the relevant question, along the lines of Kantian teleological analysis, " 'What purpose has this behavior' for the survival of this species?" Yet, according to our analysis as well as according to a review by O'Hear (2000), her answers relating origins of aesthetic to sex are "all too clearly... restricted" and represent a "wasted opportunity" to answer an important question of the biological basis for aesthetic emotion. I would also like to add that the idea that beauty is related to sex is a pervading one. When at an intermediate stage of my research I was discussing the biological and neural origins of aesthetic emotions with a leading authority in the field of cognitive and neural science, he told me that he was contemplating this problem for many years, but the relationship between beauty and sex confuses the issue and makes this matter impenetrable to scientific analysis (a private discussion). A separate discussion below is devoted to this issue because of its importance.

Carroll (2000) offers a philosophical and psychological analysis of aesthetic experience similar in some ways to the results of NMF-DL, yet opposite in conclusions. Comparing the two helps understanding the deep and subtle issues of the aesthetic theory and clarifies close but non-trivial relationships between the cognitive-mathematical theory developed here, and explanations offered previously. Carroll's analysis is multi-"dimensional" (he addresses multiple aspects of aesthetics) and deliberately careful (he discusses several approaches to controversial issues, often opposing each other). Therefore, the following comparison takes the reader through Carroll's arguments one by one and compares them to the MFT-DL analysis (for shortness, in the following **C** designates Carroll's arguments and **M** mathematical).

(1) **C** suggests that "the broadest approach... to aesthetic (of) art... encompasses any appropriate response to an artwork." A deficiency of this definition is its circularity: the art is defined through what is appropriate for art. **M** shows that aesthetic is an inherent part of *any* perception and cognition of any object or thought, and is an emotion evaluating satisfaction or dissatisfaction of KI.

(2) **C** "appropriate response" is governed by historical context of norms. **M** similarly concludes that the entire previous experience may affect every act of perception and cognition, including aesthetic perception of art. (Many authors concentrated on the historical definition of art, emphasizing that an artwork cannot be properly perceived without knowing the history of art and art criticism (Dickie 1974, Walton 1970, Danto 2001); **M** makes even a stronger statement that the entire life experience may affect every act of perception and cognition; this point will be elaborated later).

(3) **C** makes a side remark suggesting that aesthetic response to nature treats "nature as if it were art". **M** states that the basic nature of aesthetic emotion is related to KI and is the same with respect to nature and art; specifics of art ought to be analyzed within the context of dynamics of collective and individual conscious and unconscious (an example of such analysis is given later, but the comprehensive analysis will have to wait until a separate publication).

(4) **C** suggests that the "therapeutic dimension" of an artwork, that is using insights derived from an artwork for improving ones life, on the one hand is clearly appropriate, but on the other is "exactly what ... aesthetic... excludes". This illustrates fundamental difficulties with understanding Kant (1790) and aesthetics. **M** clearly distinguishes *aesthetic* emotion corresponding to the process of *cognition* of a new concept from emotions related to the processes of *using* this new concept (or even anticipating its use). The distinction is due to different instinctual and neural mechanisms involved and is not necessarily consciously perceived. Aesthetic is related to KI.

(5) **C** continues the previous line of argument to the following antinomy: "the aesthetic is... in contrast... to the cognitive" and yet the cognitive is often an important aspect of the artwork. **M** clearly specifies the relationships between aesthetic-emotional and cognitive-conceptual in processes of perception and cognition: aesthetic emotion evaluates cognitive phenomena (mental models of perception and cognition – and therefore is closely related to the cognitive), yet, the emotion is distinct from any concept, emotion evaluates not according to a rule or concept, but instinctively according to satisfaction or dissatisfaction of KI.

(6) In discussing "disinterested and sympathetic attention" **C** mentions "contemplation of an artwork for its own sake... freedom from the pressures of everyday life... We suspend questions of what is to be done personally, morally, and politically... we temporarily hang the pursuit of knowledge and truth on a hook in... the coat-check room." **M** explains that to experience the pure aesthetic emotion we do not have to suspend anything: our mind works in parallel along many pathways, and while thinking about pragmatic goals we may experience (consciously or only partly so) pure aesthetic emotions related to improving our concept-models and satisfying KI. We do not contemplate "an artwork for its own sake", but for the satisfaction of the instinct for knowledge. We definitely do not "hang the pursuit of knowledge..." because the increase of knowledge is what causes the pleasant aesthetic emotion. (This last confusion in **C** arguments is especially symptomatic and similar in its origin to the previously discussed misconception about mind as a system of logical rules, a misconception widely shared among "lay public", philosophers, and mathematicians). Whereas suspension of the pragmatic may intensify the aesthetic aspect of an experience, this is not a fundamental requirement for aesthetic experience.

(7) **C** discusses disinterested attention as directed to the artwork itself: "Are its structures unified, is it complex, what are its noteworthy expressive properties...?" **M** suggests that the specific properties of an artwork that affect us ought to be analyzed in the entire context of culture, individual mind content and its current state. And even at this level of expertise individual judgments of taste differ. The individual aesthetic judgment is only governed by the individual experience. This analysis pertains to a different level of details than the fundamental questions of the nature of "disinterested" and aesthetic.

(8) **C** discusses "sympathetic" aspect of aesthetic experience as "object-directed... surrendering to the (art)work – allowing ourselves to be guided by its structures and purposes". **M** suggests that *conscious* "surrendering" is not a necessary fundamental aspect of the aesthetic experience. There are mechanisms in the mind (conscious and unconscious) that to varying degrees enable or disable learning of new concepts, that is the aesthetic ability, at high levels of the hierarchy (obviously disabling aesthetic ability at the lower perceptual level will lead to basic incapacity and death). I prefer (in accord with the mathematical analysis) to speak about adaptivity or openness (to new experiences and ideas). The mind by evolutionary necessity is conservative. Too much openness to new experiences would have been dangerous for our ancestors. Human mind is more adaptive, more open to novelty than mind of animals. Anthropologists concluded that during the course of evolution human sexual development was modified, it is arrested at a young age and childhood prolonged so that we can learn more. Adaptivity of mind in many people is reduced after adolescence. Yet, many retain openness and adaptivity throughout life. While it is possible to develop mathematical measures of "optimal openness", it is not clear that human mind follows this optimality. Clearly, a degree of openness (sympathy) is necessary for aesthetic experience. However, an idea that more sympathy implies a better aesthetic ability is wrong. A complicated interplay of conscious and unconscious mechanisms regulates adaptivity and openness. At the higher level KI is related to the meaning and purpose of life. Openness to meaningless experiences has nothing to do with aesthetic ability.

(9) **C** discusses the potential conflict between "disinterested-ness and... sympathetic attention" related to religious, political, moral or cognitive content and purposes of artworks. **M** resolves this conflict essentially the same way as discussed in (4) and (5) above. "Sympathy" enhanced by non-aesthetic interests might predispose mind to a greater openness towards specific aesthetic experiences. Yet, sympathy could be of an aesthetic or non-aesthetic origin. Whereas the aesthetic and non-aesthetic emotions might not be clearly separated in one's consciousness, specific and different mechanisms of mind are at work. I again repeat that aesthetic is related to KI and better knowledge, and at the top of the mind hierarchy to the meaning of life.

(10) **C** resolves the conflict (9) by suggesting that aesthetic part of experience is the one that does not involve sympathy, and that aesthetic is just a part of appropriate art experience (which involves aesthetic and sympathetic experiences). **M** agrees with **C**, that an appropriate art experience is not limited to an aesthetic one. Yet **M** response is different in that first, it emphasizes that sympathy (or a part of it) might be of aesthetic origin, and second, that the aesthetic is not subsumed by the art, aesthetic experience involves non-art objects.

(11) *Evolution, sex, and the beautiful.*

**C** examines the role of evolutionary mechanisms as ultimately responsible for the aesthetic ability. This leads to a conclusion that, ultimately, aesthetic ability has specific instrumental (pragmatic) purposes. **C** identified two types of purposes: expressive and interpretative, and concludes that both are pragmatic instrumental ones. One is aimed at improving our pattern recognition ability, another at improving our intra-species communication skills, leading to cultural evolution. Both therefore are instrumental and pragmatic, and the conflict between aesthetic and pragmatic is not resolved.

    **M** analysis is threefold: (i) in agreement with **C** it concludes that in a hierarchical system, including individuals (at a lower level) and specie (at a higher level), aesthetic ability at the lower level may serve a pragmatic goal at the higher level; (ii) nevertheless, this does not deny the fundamental difference between aesthetic and pragmatic at the level of individuals: aesthetic and pragmatic goals "are implemented" through entirely different mechanisms of body-mind (KI and procreation); but this is yet not a full story: (iii) only part of our aesthetic abilities can be scientifically attributed to evolutionary-pragmatic mechanism at specie level; the top levels of the hierarchy of our mind are not subsumed within individual-specie hierarchy; first, this is suggested by what we know about hierarchical organization of our brain: the hierarchy is only approximate, interactions encompassing several levels of "hierarchy" are present in the brain; second, this is evident from our ability to formulate and pursue goals beyond the limits of specie's pragmatic interests; at the top levels of the mind hierarchy we are not satisfied with considering our ultimate goal to be the specie or gene propagation; we are searching for the "aimless" purpose, not limited by any finite aim. The ideal of beauty involves an open infinite mystery (mathematically it is related to combinatorial complexity of the choice of beautiful, which involves infinite information); Kant, analysing this issue, came with the same conclusion (1790, §§65, 80, 83). DL explains this mystery through interaction of cognition and language; but this mystery cannot be resolved, it is an actual mystery related to the meaning of life, and human lives in this mystery.

    Let me continue this discussion because some might firmly believe that evolution leads to a pragmatic explanation of our aesthetic abilities, considering such a statement an inevitable scientific conclusion of the evolutionary theory. First, this is just a belief, there is no scientific evidence that evolution can exhaustively explain the aesthetic ability. (And any attempt at predicting the future direction of science would certainly have been futile during the last 250 years). Second, consider our mathematical ability. Obviously, evolution favors some mathematical ability, say, to catch a rabbit, a hunter needs to be able to anticipate its trajectory. With what accuracy? Probably, 10% (or 0.1) would suffice, well, let us be generous, say 1% (0.01). Therefore, it might be reasonable to conclude that evolution might lead to our ability to predict events in the world of phenomena with accuracy 0.01. But how would one go about explaining our ability to predict measurements in certain physical experiments with accuracy of 0.0000000001 ? There are no scientific explanations for the origins of such ability, nor for the reason why the "real world" is mathematically predictable with such accuracy.

    (12) In Jerome vs. Charles mental experiment **C** attempts to resolve the contradiction between the pragmatic vs. "for its own sake," by considering the role of beliefs. While assuming Jerome and Charles having different beliefs, **C** maintains "they are processing the artwork in exactly the same way." **M** comments that this consideration makes no sense, it is much affected by the "logical" conception of mind discussed previously. Aesthetic experience is not a matter of conscious beliefs, but is an instinctive emotional reaction, which can be affected by conscious considerations only in an indirect ways. Separating "beliefs" from "data processing" in mind, based on what we know about the mind operations, is wrong for any purpose, and is a dangerous line of argument that leads **C** to fallacious conclusions (this Carroll's argument is also historically related to that of Dickie, criticized previously).

(13) Despite of some similarity between **C** and **M** arguments, the differences result in opposite conclusions on the most fundamental issues. **C** concludes that the only possible definition of aesthetic experience is the one of empiricism: "we can give content to the notion of aesthetic experience by merely enumerating the kind of activities and corresponding experiences." This contradicts to the basic intuition about the beautiful – everyone may just feel it. **M** concludes that aesthetic experience is a special kind of experience due to specifically aesthetic mechanisms active in the mind of everyone.

Carroll's analysis demonstrated the depth of conceptual penetration informed by many recent advances in psychology and cognitive science. Yet, subtleties of interaction between instincts, emotions, and concepts in our mind have not been clarified and lead to fundamental difficulties. I hope the above discussion helps clarify the controversial issues.

In discussing Carroll's paper, Stecker (2001) continued the mental experiment with Jerome and Charles by considering them taking up an offer from a psychologist to attend his lab and to receive subliminally "evolutionary advantages" similar to what can be derived from observing an artwork but to an even greater degree. Stecker then concluded that Charles, because he did not feel a need anymore to look at the painting did not have any aesthetic experience in the first place. This whole "mental experiment" suffers from the same "logical" conception of mind, which leads it to the wrong conclusion. After all, this psychologist lab is very much like our real life experience; while moving through life, our aesthetic taste does change, and as it evolves, we might enjoy not any longer an artwork that once was a source of genuine aesthetic experience. Mathematical theory suggests that this would happen when unconscious contents addressed by the artwork became fully conscious. This essentially dynamic nature of the interacting conscious and unconscious, which makes up the aesthetic taste, is missed by both authors.

Interaction between conscious and unconscious was missing in the above discussion of the nature of artistic experience. According to the DL analysis of mind, creativity is directed at improving (and developing new) concept-models of mind (what Kant called the faculty of understanding). Similar ideas of artistic creativity relating sensory data to novel universal conceptions have been discussed in aesthetic literature (Arnheim, 2001). The new concept-models are developed in two directions, from within, from a priori unconscious archetypes by differentiating them, connecting to the individual life experience and bringing to consciousness, and from the outside, from concept-models existing in collective conscious[3], culture, primarily in language, by connecting them to the unconscious inner instincts, through archetypes. Creativity thus involves conscious and unconscious, and occurs at the border between the two. Artistic intuitions about the self and world are internal perceptions and feelings related to vague, unconscious models in the process of connecting them to consciousness by means of artistic expression. This mathematical analysis is essentially in agreement with the account of the nature of creativity given by Jung (1934), who explained that the depth of unconscious is collective, same among all people, and that the special gift of creativity involves individual consciousness.

In the context of aesthetic creativity, the role of conscious and unconscious was explored by Croce (1922) who "identified intuitive... with... artistic fact". In correspondence with the mathematical theory and contrary to popular beliefs he emphasized first, that functioning of mind of genius is distinct from "ordinary" only in degree: "Molière... was right: 'whenever we speak, we create prose'." And, second, that both conscious and unconscious are a part of artistic experiences: "...those who claim unconsciousness as the chief quality of an artistic genius, hurl him from an eminence far above humanity to a position far below it. Intuitive or artistic genius... is always conscious." In attempting to analyze the complicated mechanism of interaction between "matter (or content) and form", however, he came with a one-sided conclusion that "The aesthetic fact... is a form, and nothing but form." This conclusion is related to the well-known Plato's view that Ideas or Forms of mind are given a priori in their final form; mathematically it is related to the idea of "logical mind." Croce arguments therefore are not consistent with his conclusion, a situation which is more typical than exceptional in the complex world of aesthetics.

Aristotle (IV BCE, Metaphysics) pointed out Plato's error: if Forms are a priori given in their final form, learning is not needed and is impossible. He discussed the difference between a priori Forms-as-potentialities (that we identify with unconscious vague models) and Forms-as-actualities (that we identify with conscious models-concepts corresponding to completed perception or cognition). Aristotle emphasized that "potentialities" are not logical, but "actualities," which emerge in the process of experience, are logical. This Aristotelian description is close to the MFT-DL theory of mind (Perlovsky 2006b, 2007c, 2010c).

I briefly discuss now the relationships between the MFT-DL theory and previously developed aesthetic-theoretical accounts of the rise and fall of Modernism and of Postmodernism. The "foremost Kantian art critic of our time (Greenberg who's) openness... and the ability to sense (Pollock's work) artistic goodness – even to proclaim its artistic greatness – at a time when this was far indeed from the received view, gave Greenberg in retrospect credentials of a kind few other critics enjoyed" (Danto 1996). According to Greenberg (1999), the main feature of Modernism is self-referentness, critical exploration of the depths of self, and abstract expressionism was the highest point in the development of art toward the ideal of disinterested beauty that explored the meanings of human self. However, his "...art criticism has become extremely problematic in an artworld almost defined by Duchamp..." (Danto, 1996) and his account does not seem to be theoretically consistent; criticizing Greenberg, De Duve (1996) and Gaiger (1999) note that while attempting to give a transcendental account of aesthetic judgment he ultimately comes to an empiricist interpretation of Kant. And according to Kuhns (2001), Greenberg "moves from enlightening openness to dogmatic closedness" and lacks "thorough grounding in the philosophical aesthetics." In view of such contradictory opinions, is it possible to provide a consistent explanation for the importance of Modernism and for its demise?

The answer to this question is "yes," both art directions satisfied opposite sides of KI. The cognitive-mathematical theory emphasizes the two aspects of the historical process of the development of mind first, that unconscious is inexhaustible in its depths extending toward the biological foundations of neural functioning that forms the material substrate of collective unconscious and unify human with animal kingdom, and second, that mind develops from fuzzy, unconscious concept-models toward more crisp and more conscious ones. The role of rational conscious mind increases in history. This process was noticed by Plato (IV BCE b), accepting for himself only the Socratic rational thinking, while mentioning the role of unconscious mystery in the creativity of oracles and poets (Perlovsky 2001). This process is not linear, it experiences delays and setbacks on large and small scales and, since the advent of scientific method in the 17$^{th}$ c., it has tremendously accelerated (Perlovsky 2007c, 2010a,b). There is an old idea that for millennia artists created unconsciously (Anichkov 1904); for instance, landscape appeared in painting much earlier than appreciation of the beauty of nature was conceptually formulated; these ideas of unconscious creativity, as discussed, are untenable. Formulating conceptually is not equivalent to being conscious. I rather think that the first cave wall painters were more conscious than their contemporaries; quite possible that "more painterly endowed" was identical with "more conscious". Yet, the conscious aspect of creativity is ever increasing; Michelangelo's "One paints, not with his hands, but with the brain" (see Gilbert 1980) could not belong to an ancient Greek artist. The reason is that the conceptual content of art tremendously increased during the 2000 years. And this process continues, "proportion" of conscious vs. unconscious in creativity is increasing; a contemporary artist expects more of his creativity than "creating beauty". A "naive" creativity became as if impossible. Especially beginning from the 19th c., an artist feels a need to combine the intuition with a rational thought (Croce 1922). Yet painting a "concept," an idea that have been already formulated in words, is not art. This pressure toward rationality, while maintaining exclusively artistic expression, is so strong that some artists feel it as oppressing and experiment with drugs to get rid of this pressure. Yet, an artist today expects "more" from his or her art than unconscious "creation of beauty." According to Danto, "artwork themselves carrying on philosophy of art" (McAdoo 2000). An artist strives toward theoretical understanding of her or his art; artistic intuition strives to unite with philosophic aesthetic thought and to become one in an artwork.

Abstract expressionism in its best achievements has penetrated deep into unconscious and revealed interesting image-less images that some perceive as aesthetically and as if magically related to archetypes of psyche (which according to Jung are space-less and time-less); its "non-representational" shapes and colors touched upon intuitions and emotions, which have never before been expressed in words or artworks. The limits of our consciousness were expanded. However, in this process ambitions of the modernist art went beyond bounds where consciousness of artists, critiques, connoisseurs, and mass-consumers of art could follow. Unconscious-as-such and unformed-as-such became the object of art.

This concentration on unconscious went against the main direction of the historical development of culture toward an increasing level of consciousness (and as well, against the great achievements of abstract art which expanded the consciousness). The "debauchery" of unconsciousness precipitated demise of abstract expressionism (and modernism in general). Postmodernist art emphasized the historic tendency toward conceptual content accessible to consciousness, towards an understandable "message", even at a price of simplification of its content. According to the MFT-DL analysis, and in correspondence with the analysis by Danto (1964) postmodernist art concentrated on the ubiquitous nature of aesthetic emotion and its fundamental basis in every act of perception and cognition. Today, being distanced by time from politics, commerce, and shockings of the moment, we can appreciate that a significant part of Minimalism, Dadaism, and Conceptualism has been just a pure aesthetic enjoyment of the availability to cognition of concepts of the simplest objects and messages. Let me repeat the previously given explanation of the Buddhist's emptiness in terms of the MFT-DL theory: an artist, like bodhisattva, "wonders at perception of emptiness in any object" (Dalai Lama, XIV, 1993) that is, concentrates on an object as a phenomenon, which concept is accessible to cognition.

Why does Danto (1999) hand "the olive branch" to Warhol's "Brillo Boxes" rather than to Duchamp's "Fountain"? An answer to this question, of course, ought to illustrate the main characteristics of any judgment of taste, its subjectivity and its objectivity. The subjective aspect is that Danto likes it better. The objective aspect according to (Danto 1999) is approximately as follows. Rejecting the previous art tradition, Duchamp concentrated on "a reaction of visual indifference with a total absence of good or bad taste" (Duchamp 1970). And eventually he conceded his non-aesthetic interests: "I threw... the urinal into their faces, and now they come and admire it for its beauty" (Duchamp 1962). Unlike Duchamp, Warhol "really was moved by the everydayness of everyday things" (Danto 1999). This pure aesthetic attitude to everydayness,

Danto sees in "Brillo Boxes" "as more philosophically challenging than the narrower, anti-aesthetic iconoclasm of Duchamp" (Danto 1999). The mathematical analysis explaining aesthetic value of any object and emphasizing this as an important aspect of postmodernism is in agreement with Danto, who values the aesthetic of pure and simple perception of any object in Warhol's artwork and identifies this value with the philosophy and art of the postmodern.

I disagree, however, with McAdoo (2000) who places Warhol closer to Breughel than to Duchamp, based on Warhol's passion for everydayness. This sounds to me as judging art according to concept. Breughel's interest in an everyday object is of totally different nature than Warhol's. In deep despair Breughel observes an everyday person. When young, he contrasts the ugliness of human "everydayness" with beauty of the surrounding landscapes. He tries to wake up people's dormant KI; he tries to build within himself a foundation for his own soul. But his attempt to reconcile with everydayness failed. When older, the despair dominates his perception, down to a total ugliness of "Blind Men". The ugliness of appearance matches the ugliness of the wretchedness of their souls. His vision is tragic, which is the opposite of both, Duchamp and Warhol. Breughel belongs to a culture, which consciousness has not yet dissociated from the tragedy of human life, but this is a different story, which I will return to later. A testimonial to the complexity of contradictions between philosophy and historicism I found in Danto (1996): "As an essentialist in philosophy... art is always the same. But as an historicist... a work of art at one time cannot be one at another..."

The enjoyment of concepts of cognition *as such* is not new to art. According to Nietzsche (1872), Homeric poets' mind has just discovered the power of concept-models that brings order to chaos of the surrounding world. This analysis by Nietzsche closely corresponds to the MFT-DL theory of the evolution of consciousness (Perlovsky 2007b,g, 2009b, 2010a) and is confirmed by an analysis of the historical changes in the conscious contents (Jaynes 1976). Homeric poets enjoyed the harmony revealed to them by their cognitive and perceptual functions. Based on this combination of Nietzsche's, Jaynes', and MFT-DL analyses, I would say that postmodernist artists attempted to return to the Apollonian consciousness of Homer. The difference however was in that for Homeric poets the just discovered conceptual was a tower, a stone wall reliably defending them from the chaotic world. Due to *conceptual* thinking, an ancient Greek felt himself a master of the world. Not so for the conceptualist artist of the $20^{th}$ c., "by the mid 1980s there seemed... no form of cultural analysis... and no ironic mode that could do more than take responsibility for its own impotence." (Altieri 2001).

The concentration on the very possibility of conscious perception of an object, in contemporary mind, is closely linked to a concentration on the inner object, which is the concept of a "real-world"-object. Here an artist, while within a relatively safe realm of consciousness, expands his or her freedom: "the attention and irony of an artist is concentrated on an internal object even when is repulsed by it" (Danto 1964). While modernism rushed toward the depth of self, the opposite tendency of postmodernism rushed with equal force towards the simplicity of the basics of aesthetics. These two opposing tendencies, sometimes combined in conscious and unconscious of an individual artist, bumped at each other, in a process that Jung called enantiodromia[4], and for a moment annihilated the art object. While intending toward conceptual, according to Danto's analysis, an artist of the 1990s "surrenders to philosophers the task of articulating its essence" (Danto 1997). Yet, Danto's conception of the end of art history does not seem convincing nor providing "a convincing account of art's essential history" (Davies 2001). It seems to me that artists of the $21^{st}$ c. will have to extricate the art from its postmodernist ambivalence between pre-Sophocles conceptualism and post-Jungian complexity of the self.

The role of the anti-aesthetic in art was much discussed and a subject of great controversies. It was at the core of initial difficulties of aesthetic theories in the 20th c. related to separating art from non-art and leading to a pessimistic idea of non-definability of art (Weitz 1956). Without discussing all the aspects of the problem, I would just mention that Dickie was among those concluding with an untenable idea that the only way to define art is empirically, according to the practice of the art world. He introduced the Institutional theory of art, first, as what was *accepted* as art (Dickie 1974), and second as what was *intended* as art (Dickie 1984). According to the first definition, as far as I can see, van Gogh was not an artist during his lifetime and became one after his death. According to the second definition, van Gogh letters are not art, and any writer's letters published posthumously without author's intent (which is unknown) are not art (or were not art before publishing, but became art after publishing? in how many copies 2, 3, 1000?). Institutional theory was much criticized for empiricism and circularity and yet, according to Matravers (2000), "there is a remarkable consensus... that something like this might be right." Reflecting on forty years of the British Journal of Aesthetics, it's Editor Lamarque (2000) noted that "from the mid-1960s, the debate (about what is art) settled broadly into two camps: ...the functions of art, and... procedures under which art was created" and "away from more traditional emphases" on properties of artwork, "away from either aesthetic responses or... aesthetic qualities." The difficulty in defining art is due to attempts to reconcile objective and subjective, and to a significant extent due to anti-aesthetic content introduced into art during the $20^{th}$ c. While anti-aesthetic was always a part of art (e.g., Goya, Brueghel the Elder, Bosch), the overwhelming influence of the ugly during the 20th c. led to "a retreat from the aesthetic". As long as "beauty" was an essential part of art, the definition of art was not felt as needed. But what is the role of "ugly" in art? Is it possible that ugly for one is beautiful for other?

How is it possible that disgusting and disharmonious could bring an aesthetic satisfaction? Upon a little thought, an answer might seem simple: an ugly scene might contain an important moral or political message. Criminal chronicle, for example, is a source of moral sweeteners, satisfying a need for compassion and a need to be convinced once more in the triumph of the moral principle. But wrong would be those seeing in these enjoyments a predominant aesthetic basis. They mostly serve other instincts, not related to improvement of the inner concept-models. A distinction between the aesthetic on the one hand and the moral, political, or other sympathies on the other, were well appreciated within the aesthetic theory, yet, in the anti-aesthetic art of the $20^{th}$ c. there was a clearly perceived (by many) an irreducible remainder of an aesthetic nature, as if an "aesthetic ugliness". But is not it an oxymoron?

A possibility of negative judgments of tastes, in particular, in Kant's aesthetics, was a subject of intense discussion in this journal and in BJA. In Thomson's opinion, ugliness cannot be reconciled with morality (Thomson, 1992); opposing arguments were presented by Guyer (1992). Recently, again Shier (1998) argued that there is no room in Kant's aesthetic for negative judgments of taste, with opposing arguments by Wenzel (1999). I find the arguments unconvincing, Wenzel's "negative purposiveness" and "disharmonious free play" are confusing, or at least not articulated clearly. This confusion is related to Kant's notion of "free play", where the word "free" indicates an absence of clear understanding of the mechanism involved[5], which is explained in the DL theory as KI. A negative judgment of taste, for example, occurs when one sees on a road a cat's body overrun by a car. A feeling of disgust, of intense aesthetic displeasure is related to this sight because it is disharmonious with our concept-models of meaning. It reminds us that spiritual meaning and purpose are limited in the material world by chance. Not the cat we pity[6], but ourselves, the dead cat reminds us of our finiteness in the material world, reminds us about as it were "insufficient basis" for concept-models of meaning and purpose of our life and contributes to destruction rather than improvement of these models. But, how is it possible, that a similarly disgusting sight of destruction can be a source of aesthetic pleasure, when rendered in oil on canvas?

Danto (1964), Levinson (1989), and other authors have discussed the fact that art perception, in principle, is related to the entire art culture and its history. I would like now to analyze deeper the mechanism by which "ugly" could contribute to the improvement of the models of meaning and ultimate purpose.

KI, as discussed previously, is a mechanism of improvement of mind's internal concept-models for better correspondences to objects and situations in the surrounding world. This process of improvement-adaptation can only go on if there is a perceived imperfection. On the contrary, if there is a perfect correspondence between the models and objects, in other words, in a state of "perfect harmony", there is no reason for internal change. A myth of "perfect harmony" appears historically when philosophers attempt to understand the ideal of beauty. A theory of beauty founded on KI reveals that a notion of "perfect harmony" corresponds to the absence of stimuli in neural networks involved in the adaptation of concept-models, in making these models clear and conscious. "Perfect harmony" is an unconscious state of the primordial unity of human and world, a state opposite to conscious creative perception that is necessary for the emergence of the feeling of beauty.

Mental concept-models emerged historically in the process of separation of human from animal kingdom, and our "humanization" is related to a slow development of better and better concepts. Therefore, new concept-models created by an artist ought to contradict in some degree to previous concepts, and consequently, satisfaction of KI is only possible in a state of broken harmony. "Theoretical harmony" should not and cannot dictate artist, which objects to pick and how to render them in artworks; because everyone has one's own experience, one's own acquired models, one's own "previous," and one's own "novel."

Aesthetic enjoyment of ugly and disharmonious is a complicated feeling related to the tragedy of human existence. A tragedy in a movie, novel, or on a theater stage could deeply move us, producing inside us at once a sad and beautiful emotion. Can we understand this emotion, if it is experienced in a purified form, if we consider it separately from a pity or compassion for the protagonist, as a pure aesthetic mixture of sadness or despair and beauty? According to the DL theory, the explanation for this complex mixture of feelings is in that we at once perceive some inadequacy of our highest models of meaning but also a possibility of improving them, and only in this combination a satisfaction of KI and therefore experience of the beautiful is possible.

Tragic art was born in Ancient Greece as a disharmony of human relationships expressed in music and text of theatrical drama, but touching not the art of visual forms. The beauty of Ancient Greek sculptures was based on an ideal of form[7]. Disharmony of visual form as an expression of human tragedy, which has sometimes been misunderstood in artworks of Goya and Breughel, only now is being accepted and justified by aesthetic thought. The nature of beauty, as perfection of the concepts of thought, of the faculty of understanding, is unified in the world of objects, in the world of thoughts, and in the world of will. In contemporary art, the disharmony of form, by reaching extreme degrees, addresses the sleeping instinct for learning. Extreme degrees of the disharmony of form (like chopping off fingers, dragging out intestines... shown on movie screens, canvas, or inspirationally described in a novel) could satisfy those with dormant aesthetic abilities. However, extreme disharmony by itself does not inspire the mass consumer of art, who loses interest in art, nor political leaders, many of whom are convinced anyway that art does not matter, nor those seeking the meaning of existence. High artworks turn to disharmony of concept-models and the world, to contradictions between concept-models, in search for new meaning, not destroying, but creating the image of beauty. Our mind, faced with the meaninglessness of the world, demands justification of being, and receives it in the beautiful, in the nature of "the aesthetic phenomenon" (Nietzsche, 1872).

## 7. Conclusion

As far as we know from Plato (IV BCEb,c), rational conscious critique of art's penetration into unconscious depths of human psyche was began by Socrates. His demand for rational understanding being an inseparable part of art, according to Nietzsche (1872) destroyed the highest achievement of Greek spirit, the Attic tragedy. This places today's stakes higher: we have to remedy what was destroyed, yet preserving the spirit of Socrates, which is the foundation of the scientific method. At the highest height of art, a victory is expected over both subjective and objective, conscious and unconscious, rational and instinctive. It is not any longer a philosopher or aesthetician demanding consciousness from an artist or poet. No. Despite Eliot's statement (Eliot 1979) that "art... performs its function... much better by indifference to (the various theories)", the artist, poet, any creative person feels that unconscious creativity is not possible any longer, feels a need to combine the intuition with a rational thought (Croce 1922; Danto 1996). Whereas centuries ago the gap between theoretical aesthetics and instinctive creativity was not even noticed by artists, today an artist is handcuffed by inadequacy of theoretical thought. Some try to overcome this oppressive gap between intuition and thought by experimenting with narcotics, however, fairy-tale narco-images do not turn into equally impressive works of art. Creative process beginning in the Dionysian chaos of the unconscious matter, in fuzzy models-archetypes making up the biological foundation of human soul has to transform the fuzzy archetypes into consciousness of crisp Apollonian forms, with necessity going through the tragedy of separation of individual from collective, subjective from objective, through the catharsis of the liberation of spirit. The tragic tension of this process is pulling stronger and stronger with the historic evolution of forms of consciousness. And searching for a possibility to combine subjective and objective an artist turns to philosophy and aesthetics, but in vain, too often contemporary thought flees the tragedy and succumbs to the 'social contract' of the 'politically correct.'

Will cognitive-mathematical analysis of mind help analyzing the pressing aesthetic questions? Would it eventually lead to a "rule" prescribing the norms of beauty? In this paper, I attempted to demonstrate that contemporary mathematics of mind is not limited to logical rules with simple "yes" or "no" and can be useful in the analysis of fundamental issues of philosophy and aesthetics. The "refusal" of mathematics to be bound by crisp logic is essential; it is a natural (for a mathematician) result of the Gödel theory that logically proved the inconsistency of crisp logic (1933). As discussed in this paper, dynamic logic connects unconscious to consciousness. In the historic development of mind, the role of crisp conscious logic is ever increasing. We exist in this dialectical paradox between vague and crisp, conscious and unconscious, and growing tensions between these opposites provide the ever-strengthening impetus for the art. This is not necessarily shocking news for a philosophically inclined one, but it is good to know that mathematics does not contradict philosophy; this knowledge may spare efforts that have periodically been devoted to doubts and confusion.

A new understanding gained with the help of cognitive-mathematical theory of mind is that beauty is a property of hierarchical adaptive systems, capable of learning and changing at many hierarchical levels. The notion of beauty is determined by contents of mental models near the top of the hierarchy, genetically inherited, transferred through culture and language, and acquired in individual experience. This hierarchy is not final, it is ever expanding, and therefore contents of representations near the top aim at this continuing expansion without limit. Therefore a computer, before coming close to the notion of beauty inherent to human mind, would have to learn and absorb human individual and cultural experience, including tactile, bodily experience, and mental models with a potentially infinite hierarchy. A "formulae of beauty", understood as a finite combination of mathematical signs which meanings are fixed axiomatically, is *impossible*, because beauty is not a finite notion. It involves a mysterious depth, infinity of perfecting our mental representations of the meaning of life (Perlovsky 2010g).

Mathematical analysis of the nature of beauty reveals the historical dialectic of the a priori and adaptive, penetrating into the interaction of objective and subjective. It leads to the understanding of the increasing role of rational and conscious "apollonian" in beautiful in interaction with unconscious and instinctual "Dionysian", as a part of the general historic process of the evolution of mind and consciousness. The development of consciousness combining such diverse principles, individuation and synthesis, subjective and objective, involves a striving toward the infinite, a "sweep of the wings of yearning", the process as tragic as history itself. It explains the nature of disharmony in the beauty not as ugly and disgusting as such, but as a mechanism of creating new meanings, mechanism of understanding the infinite "aimless" purpose of our existence – improving our internal image of Self.

De Duve, T. 1996, Clement Greenberg Between the Lines, (Paris: Editions Dis Voir, tr. B., Holmes).
Deming, R., Perlovsky, L.I. (2005). *A Mathematical Theory for Learning, and its Application to Time-varying Computed Tomography.* New Math. and Natural Computation. **1**(1), 147-171.
Deming, R.W. and Perlovsky, L.I. (2007). Concurrent multi-target localization, data association, and navigation for a swarm of flying sensors, *Information Fusion,* **8**, 316-330.
Dickie, G. 1984, The Art Circle, Haven, New York, NY.Dickie, G. 1964, The Myth of the Aesthetic Attitude, In Ph. Alperson, ed. The Philosophy of the Visual Art, Oxford. Univ. Press, NY, 1992
Dickie, G. 1974, What Is Art: An Institutional Analysis, in Ph.Alperson, ed., The Philosophy of the Visual Art, Oxford. Univ. Press, NY, 1992.
Dickie, G. 2004, Reading Sibley, Brit J Aesthetics, 44: 408-412
Duchamp, M. In 1970, H.H.Arnason, History of Modern Art, Prentice Hall, England Cliffs, N.J., p.229.
Duchamp, M. In 1962, H.Richter, Dada: Art and Anti-Art, Thames and Hudson, 1966, London, p.207.
Eliot, T.S. 1979, The Function of Criticism, in Selected Prose of T.S. Eliot, ed. F. Kermode, Faber, London.
Fontanari, J.F. and Perlovsky, L.I. (2004). *Solvable null model for the distribution of word frequencies.* Physical Review **E 70**, 042901 (2004).
Fontanari, J.F. and Perlovsky, L.I. (2007). Evolving Compositionality in Evolutionary Language Games. *IEEE Transactions on Evolutionary Computations*, **11(6)**, 758-769; doi:10.1109/TEVC.2007.892763
Fontanari, J.F. and Perlovsky, L.I. (2008a). How language can help discrimination in the Neural Modeling Fields framework. *Neural Networks*, **21**(2-3), 250–256
Fontanari, J.F. and Perlovsky, L.I. (2008b). *A game theoretical approach to the evolution of structured communication codes*, Theory in Biosciences, **127**, 205-214, http://dx.doi.org/10.1007/s12064-008-0024-1.
Fontanari, F.J., Tikhanoff, V., Cangelosi, A., Ilin, R., and Perlovsky, L.I.. (2009). Cross-situational learning of object–word mapping using Neural Modeling Fields. *Neural Networks, 22 (5-6),* 579-585
Gaiger, J. 1999, Constraints and Conventions: Kant and Greenberg on Aesthetic Judgment, British Journal of Aesthetics, 39(4), p.376-391.
Gaut, B. 2000, 'Art' as a Cluster Concept, in Noël Carroll (ed.), Theories of Art Today. Madison: University of Wisconsin Press, pp. 25-44
Gaut, B. 2005, The Cluster Account of Art Defended, Brit J Aesthetics, 45: 273-288
Gödel, K. 1986, Kurt Gödel collected works, ed. S.Feferman at al, Oxford University Press.
Greenberg, C. 1999, Homemade Esthetics: Observations on Art and Taste, Oxford: Oxford Univ. Press.
Grossberg, S. (1982). Studies of Mind and Brain. D.Reidel Publishing Co., Dordrecht, Holland.
Grossberg, S. & Levine, D.S. (1987). Neural dynamics of attentionally modulated Pavlovian conditioning: blocking, inter-stimulus interval, and secondary reinforcement. Psychobiology, **15**(3), pp.195-240.
Guyau, J.M. 1884/2010, Les problemes d'esthetique contemporaine. Nabu Press (French).
Guyer, P. 1992, Thomson's Problem with Kant: A Comment on "Kant's Problems with Ugliness", JAAC, **50**(4), pp.317-319.
Guyer. P. Ed. 1992.The Cambridge companion to Kant, Cambridge, UK: Cambridge University Press.
Guyer, P. 1997, Kant and the Claims of Taste, Cambridge U.P., Cambridge.
Hartmann, v. E. 1904, In V. Soloviov, Aesthetics, Encyclopedic Dictionary, ed. F.Brokhause & I.Efron, St.Peterburg, Russia, (Russian). **(find on Amazon)**
Ilin, R. & Perlovsky, L.I. (2010). Cognitively Inspired Neural Network for Recognition of Situations. *International Journal of Natural Computing Research,* 1(1), 36-55.
Jaynes, J. 1976, *The Origin of Consciousness in the Breakdown of the Bicameral Mind.* Houghton Mifflin Co., Boston, MA.
Jung, C.G., 1921, *Psychological Types.* In the Collected Works, v.6, Bollingen Series XX, 1971, Princeton University Press, Princeton, NJ.
Jung, C.G. 1934, *Archetypes of the Collective Unconscious.* In the Collected Works, v.9,II, Bollingen Series XX, 1969, Princeton University Press, Princeton, NJ.
Kant, I. 1790, The Critique of Judgment, tr. J.H.Bernard, Prometheus Books, Amherst, NY.
Kemp, G. 1999, The Aesthetic Attitude, BJA, 39(4), pp.392-399.
Kovalerchuk, B. & Perlovsky, L.I. (2008). Dynamic Logic of Phenomena and Cognition. IJCNN 2008, Hong Kong, 3530-3537.
Kozma, R., Puljic, M., and Perlovsky, L. (2009). *Modeling goal-oriented decision making through cognitive phase transitions*, New Mathematics and Natural Computation, 5(1), 143-157.
Kuhns, R. 2001, Book Review, JAAC, **59**(1), pp.97-99.
Lamarque, P. 2000, The British Journal of Aesthetics: Forty Years On, BJA, **40**(1), pp.1-20.
25                    25 3:40 PM; 12/10/10

Perlovsky, L.I. (2010b). *Intersections of Mathematical, Cognitive, and Aesthetic Theories of Mind*, Psychology of Aesthetics, Creativity, and the Arts, 4(1), 11-17. doi: 10.1037/a0018147.

Perlovsky, L.I. (2010c). Neural Mechanisms of the Mind, Aristotle, Zadeh, & fMRI, *IEEE Trans. Neural Networks*, 21(5), 718-33.

Perlovsky, L.I. (2010d). The Mind is not a Kludge, *Skeptic*, 15(3), 51-55

Perlovsky, L.I. & Ilin, R. (2010a). Neurally and Mathematically Motivated Architecture for Language and Thought. *Special Issue "Brain and Language Architectures: Where We are Now?" The Open Neuroimaging Journal,* 4, 70-80. http://www.bentham.org/open/tonij/openaccess2.htm

Perlovsky L.I. & Ilin, R. (2010b). Grounded symbols in the brain. IEEE World Congress on Computational Intelligence (WCCI'10), Barcelona, Spain; arXiv; WebmedCentral.

Perlovsky, L.I, Bonniot-Cabanac, M.-C., & Cabanac, M. (2010). Curiosity and pleasure, arXive 1010.3009.

Perlovsky, L.I. (2010e). Science and Religion: *Scientific Understanding of Emotions of Religiously Sublime,* arXive.

Perlovsky, L.I. (2010f). Jihadism and Grammars. Comment to "Lost in Translation," Wall Street Journal, June 27, http://online.wsj.com/community/leonid-perlovsky/activity

Perlovsky, L.I. (2010g). Joint Acquisition of Language and Cognition; WebmedCentral BRAIN;1(10):WMC00994; http://www.webmedcentral.com/article_view/994

Perlovsky, L.I. (2010h). Freedom of will, arXive.

Perlovsky, L.I. (2010). Dynamic logic; http://www.scitopics.com/Dynamic_logic.html

Perlovsky, L.I. (2010). Dynamic logic, computational complexity, engineering and mathematical breakthroughs; http://www.scitopics.com/Dynamic_logic_computational_complexity_engineering_and_mathematical_breakthroughs.html

Perlovsky, L.I. (2010). Dynamic Logic. Cognitive Breakthroughs; http://www.scitopics.com/Dynamic_Logic_Cognitive_Breakthroughs.html

Perlovsky, L.I. (2010). Mind mechanisms: concepts, emotions, instincts, imagination, intuition, beautiful, spiritually sublime; http://www.scitopics.com/Mind_mechanisms_concepts_emotions_instincts_imagination_intuition_beautiful_spiritually_sublime.html

Perlovsky, L.I. (2010). Language and Cognition. Interaction Mechanisms; http://www.scitopics.com/Language_and_Cognition_Interaction_Mechanisms.html

Perlovsky, L.I. (2010). Languages and Cultures: Emotional Sapir-Whorf Hypothesis (ESWH); http://www.scitopics.com/Languages_and_Cultures_Emotional_Sapir_Whorf_Hypothesis_ESWH.html

Perlovsky, L.I. (2010). Jihad and Arabic Language. How Cognitive Science Can Help Us Understand the War on Terror; http://www.scitopics.com/Jihad_and_Arabic_Language_How_Cognitive_Science_Can_Help_Us_Understand_the_War_on_Terror.html

Perlovsky, L.I. (2010). Music and Emotions. Functions, Origins, Evolution; http://www.scitopics.com/Music_and_Emotions_Functions_Origins_Evolution.html

Perlovsky, L.I. (2010). Beauty and art. Cognitive function and evolution; http://www.scitopics.com/Beauty_and_art_Cognitive_function_and_evolution.html

Perlovsky, L.I. (2010). New "Arrow of Time": Evolution increases knowledge; http://www.scitopics.com/New_Arrow_of_Time_Evolution_increases_knowledge.html

Perlovsky, L.I. (2010). Science and Religion: New cognitive findings bridge the fundamental gap; http://www.scitopics.com/Science_and_Religion_New_cognitive_findings_bridge_the_fundamental_gap.html

Plato. (IV BCa). *Cratylus*. Tr. C.D.C.Reeve in Plato, ed. L.Cooper, Oxford University Press, New York, NY.

Plato. (IV BCb). *Parmenides*. Trans. in Plato, L. Cooper, Oxford University Press, New York, NY.

Plato. (IV BCc). Republic. Trans. in Plato, L. Cooper, Oxford University Press, New York, NY.

Putnam, H. 1988, Representation and Reality, MIT Press, Cambridge, MA.

Schiller, F. 1895/1993, In Essays, German Library, v. 17, ed. W.Hinderer and D.Dahlstrom, Continuum Pub Group.

Scruton, R. 1974, Art and Imagination, Routledge, London.

Scruton, R. 2007, In Search of the Aesthetic, The British Journal of Aesthetics, 47(3):232-250

Shier, D. 1998, Why Kant Finds Nothing Ugly, BJA, **38**(4), pp.412-418.

Spencer, H. 1891, In Essays, Scientific, Political and Speculative, Williams & Norgate, London.

Stecker, R. 2001, Only Jerome: a Reply to Noël Carroll, BJA, **41**(1), pp.76-80.

Stolnitz, J. 1960, The Aesthetic Attitude, In Ph. Alperson, ed. The Philosophy of the Visual Art, Oxford. Univ. Press, NY, 1992.

Thomson, G. 1992, Kant's problem with ugliness, JAAC, **50**(2), pp.107-115.

---

[1] Journal of Aesthrtics and Art Criticism, British Journal of Aesthtics, Psychology of Aesthetics, Creativity and the Arts.

[2] Absence of dynamics in Kantian theory is almost unnoticed, among exceptions is N.Carroll, 2001, Modernity and the Plasticity of Perception, JAAC, **59**(1), pp.11-17.

[3] The notion of *collective conscious* introduced here refers to the entire set of conceptual knowledge existing in culture, in the entire way of life, and primarily in language; Jung introduced *collective unconscious* emphasizing that unconscious contents of one's mind are not under the control of individual self, are not individual in nature, but similar among all people. Both collective conscious and unconscious do not "belong" to an individual. Creativity is a process of connecting them, this process is a symbol-process which makes both conscious and unconscious (to some extent) available to individual (Jung 1934).

[4] Enantiodromia is a condition of mind identified by Jung (1921), in which conscious and unconscious tendencies oppose each other, leading to a demise of self.

[5] As distinct from the fundamental "freedom", which is the a priori principle of practical reason.

[6] I am not suggesting that pitying the cat is bad or wrong, I am emphasizing a deeper nature of the feeling of pity.

[7] "Laocoon" was created late in Ancient Greek history; it is an exception that we do not discuss here for lack of space.